\documentclass[journal]{IEEEtran}
\usepackage{cite}
\usepackage{amsmath,amssymb,amsfonts}
\usepackage{hyperref}
\usepackage{graphicx}
\usepackage{textcomp}
\usepackage{bm}
\usepackage{relsize}
\usepackage{amsmath}
\usepackage{amsmath, amsthm, amssymb}
\usepackage{float}
\usepackage{graphicx}
\usepackage{epstopdf}
\usepackage{lipsum}
\usepackage{caption}
\usepackage{algorithm}
\usepackage{algpseudocode}
\usepackage{verbatim}
\usepackage{calligra}
\usepackage[font={footnotesize}]{caption}
\usepackage{subfigure}
\usepackage{xcolor}
\usepackage{mathtools}
\usepackage{stackengine}
\usepackage[scr=boondoxo,
scrscaled=1.15,]{mathalfa}
\usepackage{bbm}
\usepackage{adforn}
\usepackage{acronym} \input{acronym}
\usepackage{tikz}
\definecolor{bg}{RGB}{237,239,223}
\definecolor{frame}{RGB}{104, 85, 95}

\DeclareMathAlphabet{\mathpzc}{OT1}{pzc}{m}{it}
\DeclareMathAlphabet{\mathcalligra}{T1}{calligra}{m}{n}

\begin{document}

\title{\textsc{Enwar}: A RAG-empowered Multi-Modal LLM Framework for Wireless Environment Perception}

\author{
Ahmad M. Nazar, Abdulkadir Celik, Mohamed Y. Selim, Asmaa Abdallah, Daji Qiao, Ahmed M. Eltawil
\vspace{-0.75cm}
}
\maketitle

\begin{abstract}
\Acp{LLM} hold significant promise in advancing network management and orchestration in 6G and beyond networks. However, existing LLMs are limited in domain-specific knowledge and their ability to handle multi-modal sensory data, which is critical for real-time situational awareness in dynamic wireless environments. This paper addresses this gap by introducing \textsc{Enwar}\footnote{Enwar is a common name in Turkic and Arabic cultures, meaning more enlightened, insightful, and intellectual; herein referring to a multi-modal \ac{LLM} providing deep situational and contextual insights into the environment.}, an \underline{\textsc{EN}}vironment-a\underline{\textsc{WAR}}e retrieval augmented generation-empowered multi-modal \ac{LLM} framework. \textsc{Enwar} seamlessly integrates multi-modal sensory inputs to perceive, interpret, and cognitively process complex wireless environments to provide human-interpretable situational awareness. \textsc{Enwar} is evaluated on the  GPS, LiDAR, and camera modality combinations of DeepSense6G dataset with state-of-the-art \acp{LLM} such as Mistral-7b/8x7b and LLaMa3.1-8/70/405b. Compared to general and often superficial environmental descriptions of these vanilla LLMs, \textsc{Enwar} delivers richer spatial analysis, accurately identifies positions, analyzes obstacles, and assesses line-of-sight between vehicles. Results show that \textsc{Enwar} achieves key performance indicators of up to 70\% relevancy, 55\% context recall, 80\% correctness, and 86\% faithfulness, demonstrating its efficacy in multi-modal perception and interpretation. 
\end{abstract}
\IEEEpeerreviewmaketitle

\section*{\textbf{Introduction}}

\begin{figure*}
    \centering
    \includegraphics[width= 1.0 \textwidth]{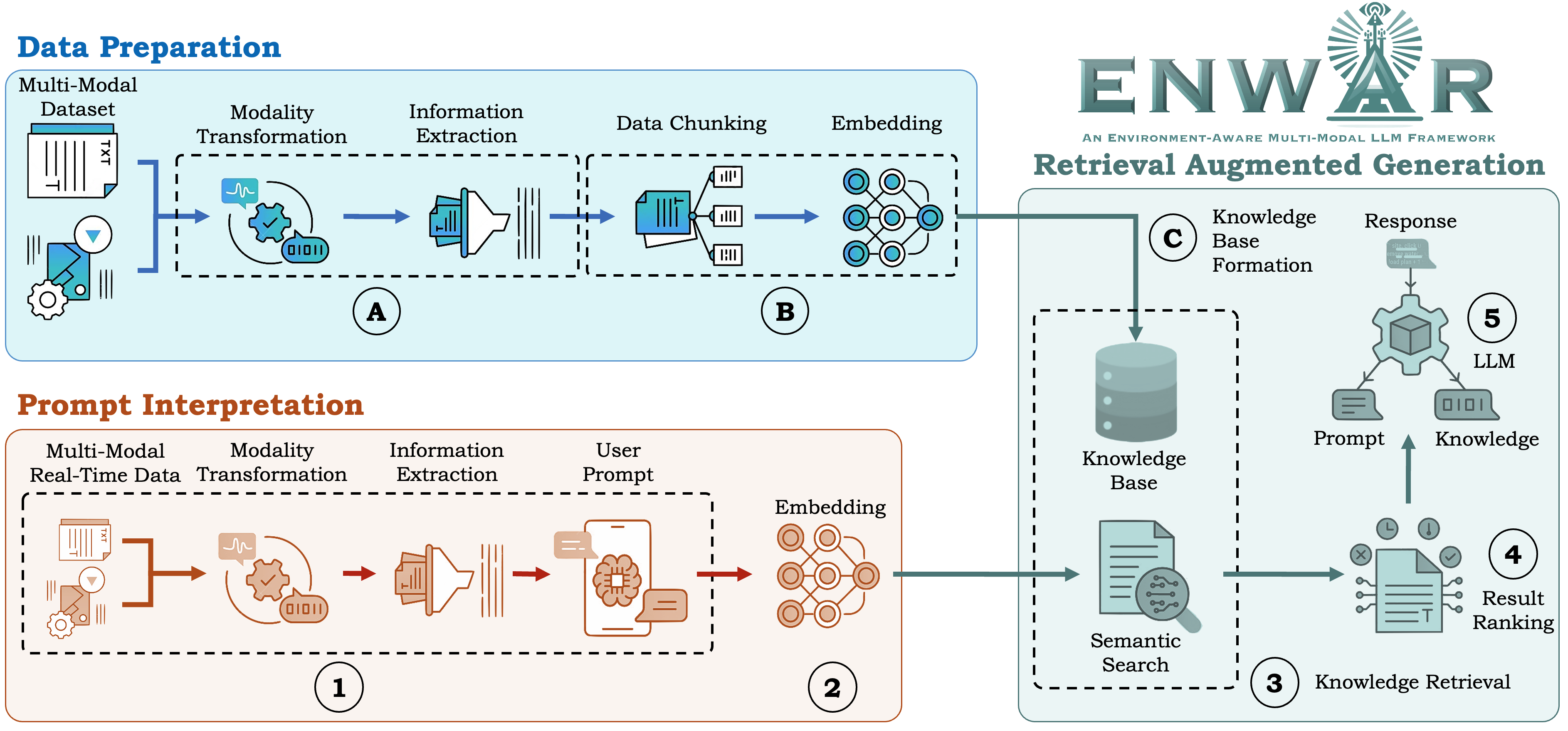}
    \caption{\textsc{Enwar} workflows: multi-modal RAG formation (Steps A-C); and  prompt interpretation, knowledge retrieval, and response generation (Steps 1-5).}
    \label{workflow}
\end{figure*}

\IEEEPARstart{G}{enerative} \ac{AI}, with its unparalleled capability to generate, synthesize, and adapt data, is poised to play a pivotal role in the evolution of 6G and beyond networks \cite{celikGenAI}. Among various generative models, \acp{LLM} have proven to be the most revolutionary, fundamentally changing how machines understand and produce human language. Built on sophisticated generative transformers and driven by attention mechanisms, \acp{LLM} leverage vast pretraining on diverse datasets to excel in tasks such as natural language processing, decision support, and beyond \cite{Vaswani2017}. \acp{LLM}' adaptability and scalability are particularly well-suited for complex systems operating in dynamic environments, making them valuable assets for advancing \ac{AI}-native wireless systems and enhancing the cognitive capabilities of next-generation networks. Hence, they have the potential to revolutionize decision-making, resource management, and intelligent optimization of 6G networks, eventually paving the way for \ac{ZSM}.

However, the technical demands of next-generation networks differ greatly from legacy generations. Future networks are expected to operate with massive antenna arrays at significantly higher frequencies, wherein wireless channels become less probabilistic and more deterministic and exhibits geometric propagation characteristics. This shift introduces daunting mobility challenges such as tracking narrow beams, blockage mitigation through timely handover management, and seamless service migration. Sensing functionalities and environmental awareness are essential for \ac{ZSM} to effectively navigate this new terrain. 

In this context, multi-modal \ac{ISAC} represent coherent fusion of disparate data streams from various sensors (e.g., LiDARs, radars, cameras, GPS, etc.), unlocking critical capabilities such as environment mapping, object/human detection and classification, urban planning, localization, and tracking. These sensing functionalities collectively lay the foundations of \acp{DT}; a dynamic and near real-time virtual replica of 6G networks, providing contextual and site-specific insights into the spatio-temporal characteristics of wireless environment \cite{Alkhateeb2023DT}. \acp{DT} are crucial in optimizing network performance, enabling real-time decision-making, and enhancing overall situational awareness, making it an integral component of the future telecom ecosystem.

Nonetheless, \acp{LLM} predominantly operate in a text-based modality, which restricts their ability to interact with multi-modal sensory data—a critical need in situation-aware networks where real-world comprehension goes beyond textual inputs. Moreover, \acp{LLM}' possession of general knowledge over a massive training data corpus often falls short on domain-specific tasks and contexts. This limitation arises from their fundamental reliance on probabilistic pattern recognition rather than true understanding or reasoning \cite{Bender2021}. To overcome these deficiencies, \ac{RAG} frameworks have emerged as a potential solution to enhance the generative process by integrating external knowledge retrieval, allowing \acp{LLM} to tap into domain-specific databases or real-time knowledge sources. This augmentation provides more accurate, contextually relevant responses, bridging the gap between generic \acp{LLM} and specialized 6G network needs. While \ac{RAG} addresses the challenge of domain expertise, it does not entirely resolve the critical need for multi-modal \ac{ISAC} to realize situational-aware 6G networks.


The integration of \acp{LLM} into wireless networks has been initially explored in \cite{2,3}, which primarily focus on textual data and overlook \ac{RAG} capabilities, limiting their application to telecom chatbots. WirelessLLM demonstrates how domain-specific knowledge can be incorporated into \acp{LLM} to enhance performance in tasks such as spectrum sensing and protocol understanding \cite{wirelessLLM}. Similarly, Xu et al. propose a framework for edge \acp{LLM} that are divided into perception, grounding, and alignment modules to optimize 6G-related tasks \cite{6gllm}. While the potential of \acp{MLLM} has been discussed in \cite{4, hashash2024}, these vision papers lack comprehensive case studies and proof-of-concept implementations to substantiate \acp{MLLM}' effectiveness in real-world multi-modal environments. We address this gap in the wireless literature by introducing \textsc{Enwar}
, an \underline{\textsc{En}}vironment-a\underline{\textsc{war}}e \ac{RAG}-empowered \ac{MLLM} framework that leverages multi-modal sensory data to perceive, interpret, and cognitively process complex wireless environments. \textsc{Enwar}'s  human-interpretable situational awareness is crucial for both sensing and communication applications, where real-time environmental perception can significantly enhance system performance and reliability. 

In the following sections, we first outline the workflow of \textsc{Enwar} and introduce \acp{KPI}, namely answer relevancy, context recall, correctness score, and faithfulness. \textsc{Enwar}'s performance is evaluated across Mistral-7b/8x7b and Llama3.1-8/70/405b models on GPS, LiDAR, and camera modalities of vehicle-to-vehicle scenarios in the DeepSense6G dataset \cite{deepsense4}. While  Vanilla \acp{LLM} provide a general and often superficial environment descriptions, \textsc{Enwar} delivers contextually rich analysis of spatial dynamics by accurately identifying the positions and distances of entities (vehicles, cyclists, pedestrians, etc.), analyzing potential obstacles, and assessing line-of-sight between communicating vehicles. Numerical results compare various modality combinations across different \ac{LLM} versions and demonstrates that \textsc{Enwar} achieves up to 70\% relevancy, 55\% context recall, 80\% correctness, and  86\% faithfullness. The paper concludes with an exploration of future research directions and the potential of \textsc{Enwar} to advance multi-modal perception and cognition in wireless systems.


\section*{\textbf{An Overview of \textsc{Enwar} Framework}}
As illustrated in Fig. \ref{workflow}, \textsc{Enwar} is comprised of two primary workflow pipelines: 1) multi-modal \ac{RAG} formation (Steps A-C) and 2) prompt interpretation, response generation, knowledge retrieval, and response generation (Steps 1-5); which are described in the following sections along with \acp{KPI}. 

\subsection*{\hspace{7 pt} \textbf{Multi-Modal RAG Formation}}

\subsubsection*{\hspace{-5pt} \textcircled{\footnotesize \textbf{A}} \textbf{Dataset Preprocessing and Modality Transformation}}
\textsc{Enwar} is designed to seamlessly accommodate diverse sensor modalities by preprocessing and transforming them into a unified textual format that can be effectively processed by \acp{LLM}. For instance, GPS data undergoes transformation from raw spatial coordinates into textual descriptions that provide insights such as relative distances, directional bearings, and movement patterns, offering a richer contextual understanding of spatial relationships.

Visual data are processed through an image-to-text conversion model that extracts key visual elements and translates them into \ac{LLM} interpretable natural language descriptions. The use of instructional prompts ensure that the generated textual outputs are contextually relevant and sufficiently detailed to accurately represent the visual information.

Point cloud data from LiDARs, another complex modality, is processed by feature extraction models (e.g., ResNet) to identify salient environmental elements. Object detection and classification systems are then employed to recognize key entities (e.g., pedestrians, vehicles), which are subsequently converted into textual descriptions.

The final step in the preprocessing pipeline involves synthesizing the transformed data from all modalities into a unified textual representation. By consolidating various sensory data into a textual format (e.g, JSON format), \textsc{Enwar} ensures that \acp{LLM} can cohesively process and interpret multi-modal inputs, enhancing the model's ability to generate contextually aware and reliable outputs. This synthesis is pivotal for enabling the framework to make informed decisions in complex environments.

\subsubsection*{\textcircled{\footnotesize \textbf{B}} \textbf{Text Chunking and Embedding}} 
\textsc{Enwar}'s next critical step is to segment the sensory data into manageable chunks and convert these chunks into numerical embeddings. In this way, \acp{LLM} can efficiently process and interpret textual information, especially when handling large datasets from diverse sensor modalities.

Chunking involves breaking down the preprocessed text into smaller, contextually coherent units. This is essential as \acp{LLM} have token limits, meaning that excessively large text inputs cannot be processed effectively. Segmentation ensures that the model can focus on relevant parts of the data without losing contextual integrity. For instance, GPS data may be chunked based on time intervals or location changes, while visual and point cloud descriptions could be divided based on objects detected or spatial regions.

Once the data is chunked, it is passed to a \ac{GTE} model to convert each chunk into a dense vectorized format—a numerical representation of the text that captures its semantic content. These embeddings serve as a structured and machine-readable format that encodes the underlying meaning of the text. In other words, vectorization enables \acp{LLM} to tokenize and process the data, establishing relationships between different chunks based on their semantic similarity.

\subsubsection*{\textcircled{\footnotesize \textbf{C}}  \textbf{Domain-Specific Knowledge Base Generation}}

\textsc{Enwar}’s ability to deliver precise and context-aware responses is largely dependent on its robust domain-specific knowledge base. By constructing a knowledge base comprising of embeddings generated from a variety of sensor modalities, \textsc{Enwar} ensures that the system is equipped with contextually rich and diverse information about the environment. To ensure optimal performance, the knowledge base is indexed in a way that allows the \ac{RAG} framework to retrieve relevant data efficiently, which is explained in the following sections. This structured knowledge base enables real-time decision-making and ensures that \textsc{Enwar} remains adaptable and responsive to a wide range of scenarios, enhancing its performance in dynamic and complex wireless environments.


\subsection*{\hspace{7 pt}  \textbf{Prompt Interpretation and Response Generation}}

\subsubsection*{\textcircled{\footnotesize \textbf{1}} \textbf{Prompt Preprocessing and Modality Transformation}} 
This step closely mirrors the procedures in Step-\textcircled{\footnotesize A}: the user prompt is preprocessed by transforming its components and any real-time multi-modal sensory data into a unified, standardized format suitable for \acp{LLM}. This ensures the prompt is properly aligned with the knowledge base, allowing for seamless interaction with the model's retrieval mechanisms.

\subsubsection*{\textcircled{\footnotesize \textbf{2}} \textbf{Prompt Text Embedding}} 
Similarly, this step follows the procedures in Step-\textcircled{\footnotesize B}: the preprocessed prompt is converted into numerical embeddings, ensuring that it can be efficiently processed and compared to the vectorized data in the knowledge base. This transformation facilitates accurate retrieval of relevant information, streamlining the prompt’s interaction with the model's generative components.

\subsubsection*{\textcircled{\footnotesize \textbf{3}} \textbf{Semantic Search and Knowledge Retrieval}} 
Once the user prompt has been transformed into embeddings, \textsc{Enwar} performs semantic search to retrieve the most relevant information from its domain-specific knowledge base. This process identifies entries that closely match the prompt by calculating the semantic similarity between the prompt and the embedded data in the knowledge base. As detailed next, the top-ranked results, which are contextually aligned with the prompt, are then selected for further processing.


\subsubsection*{\textcircled{\footnotesize \textbf{4}} \textbf{Result Ranking}}
{\textsc{Enwar} ensures relevance by ranking results according to their section headers, prioritizing the most contextually appropriate portions of the knowledge base. This refined search mechanism optimizes retrieval by focusing on the most pertinent content. Since some contexts may have similar vectorized embeddings, \textsc{Enwar} concentrates on the top-$p$ percentile to effectively filter out less relevant data, with $p=95$ used throughout the system to anchor the retrieval process in the highest-ranking results. This approach enhances both the precision and relevance of the retrieved information, resulting in more accurate and contextually appropriate responses.}


\subsubsection*{\textcircled{\footnotesize \textbf{5}} \textbf{Response Generation}}
Once the top-ranked results from the semantic search are identified, they provide the essential context for the \ac{LLM} to generate a coherent and contextually appropriate response. These results serve as the foundation upon which the \ac{LLM} builds its output, ensuring that the generated response is both accurate and relevant to the user’s prompt.

The \ac{LLM} processes the vectorized embedding of the user prompt along with the retrieved context from the knowledge base. It integrates information from multiple sources, such as GPS coordinates, LiDAR data, and visual descriptions, to construct a detailed representation of the environment. This may involve detecting vehicles, their locations, and describing physical aspects of the surroundings in relation to the prompt.
{To further enhance the generation process, \textsc{Enwar} employs top-$p$ sampling to strike a balance between accuracy and diversity in responses, effectively filtering out irrelevant outputs while maintaining contextual richness.}

Beyond simple description, the \ac{LLM} can infer interactions between various elements in the environment. For instance, using GPS data and environmental context, the model might predict how vehicles interact or how the physical surroundings could influence those interactions. 
Furthermore, the \ac{LLM} ensures that the response aligns with the specific instructions or tasks provided by the user. In the context of \textsc{Enwar}, this means delivering comprehensive environmental awareness by describing key entities, their locations, and how they may interact within the given environment. By synthesizing all relevant data and ensuring it is grounded in the retrieved context, the \ac{LLM} generates a detailed and actionable response, providing insights necessary to make informed decisions, particularly in dynamic and complex scenarios.


\subsection*{\hspace{7pt}  \textbf{Key Performance Indicators}} 

Evaluating the performance of \textsc{Enwar} requires assessing its output based on both general benchmarks and domain-specific metrics. Standard benchmarks such as the \ac{GLUE} and \ac{MMLU} offer a broad assessment of an \ac{LLM}’s capabilities across various metrics such as answer relevancy, factual correctness, and hallucinations avoidance. However, \ac{RAG}-based systems require a more targeted evaluation due to their reliance on domain--specific contexts, and in our case, specifically tailored multi-modal data. Following RAGAS framework [\url{https://docs.ragas.io}], we ensure a comprehensive and accurate evaluation of \textsc{Enwar}'s performance through following \acp{KPI}:

{\hspace{-10 pt} {\large \adforn{3}} \textbf{\textit{Answer Relevancy}}} {(AR) measures how well \textsc{Enwar}'s responses align with the user's prompt and context. Denoting $E_{p_i}$ and $E_{t_i}$ as the embedding vectors of $i^{th}$ generated prompt and the relevant ground truth of the data sample, respectively; cosine/semantic similarity, $-1 \leq \cos\left(\vec{E}_{p_i}, \vec{E}_{t_i}\right) \leq 1$, measures how semantically similar two texts are based on their vector representations, i.e., 1: perfectly similar (aligned), 0: no similarity, -1: completely dissimilar (opposite). Accordingly, AR is evaluated by calculating the average cosine similarity as follows
\begin{equation}
\text{AR}=\frac{1}{N} \sum_{i=1}^N \cos\left(\vec{E}_{p_i}, \vec{E}_{t_i}\right) =\frac{1}{N} \sum_{i=1}^N \frac{\vec{E}_{p_i} \cdot \vec{E}_{t_i}}{\left\|\vec{E}_{p_i}\right\|_2\left\|\vec{E}_{t_i}\right\|_2}.
\end{equation}
}
\hspace{-10 pt} {\large \adforn{3}} \textbf{\textit{Context Recall}} assesses the degree to which \textsc{Enwar}’s retrieved context aligns with the ground truth. It evaluates whether the system correctly recalls information from the knowledge base that is relevant to the prompt and verifies how much of the response can be attributed to the correct context. It is calculated by normalizing the alignment extent of the retrieved contexts within the ground truth with the number of sentences in the ground truth.

{\hspace{-10 pt} {\large \adforn{3}} \textbf{\textit{Correctness Score}}} {measures the factual correctness and semantic similarity of the generated responses. Denoting the embedding vector of $i^{th}$ generated answer by $\vec{E}_{a_i}$ and $F_1$ score as a metric of factual correctness, the overall correctness score is given by 
\begin{equation}
\text{Correctness}=\omega \cos\left(\vec{E}_{a_i}, \vec{E}_{t_i}\right) + (1-\omega) F_1,
\end{equation}
where weighting parameter $0 \leq \omega \leq 1$\textemdash the RAGAS sets $\omega=0.25$ as default\textemdash ensures that response assessments are both factually accurate and contextually appropriate.}

{\hspace{-10 pt} {\large \adforn{3}} \textbf{\textit{Faithfulness}}} evaluates the consistency of the generated answers with the retrieved context. A response is considered faithful if all claims align with the retrieved data, ensuring the output does not contain unsupported or fabricated information. Faithfulness checks whether the claims made in the output can be logically deduced from the given context and is given by
\begin{equation}
    \text{Faithfulness}=\frac{|N_{G_c}|}{|N_C|},
\end{equation}
where $N_{G_c}$ is the number of claims in the generated answer that can be inferred from the given context and $N_C$ is the total number of claims in the generated answer.
\begin{figure*}
    \centering
    \includegraphics[width=\textwidth]{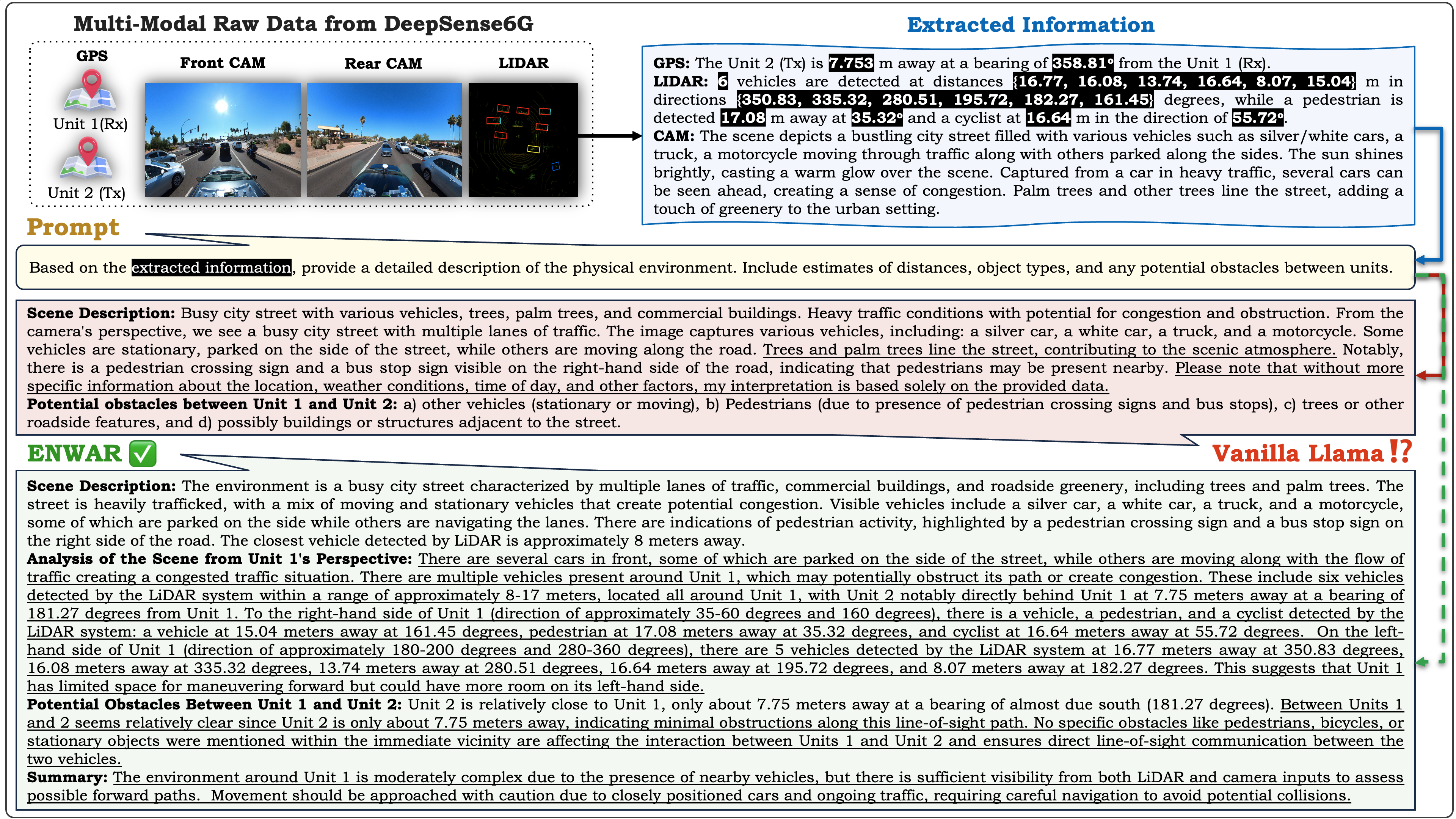}
    \caption{Illustration of the case study scene with raw data, extracted information, generated prompt, and responses from Vanilla Llama and \textsc{Enwar}.}
    \label{fig:sample_answer}
\end{figure*}


\section*{\textbf{\textsc{Enwar} Setup Breakdown and A Case Study}}
This section provides a detailed breakdown of the \textsc{Enwar} setup and offers a qualitative performance comparison with vanilla \acp{LLM}, highlighting the advantages of integrating multi-modal data and knowledge retrieval within \textsc{Enwar}.

\subsection*{\hspace{7pt} \textbf{A Breakdon of \textsc{Enwar} Setup}}

\subsubsection*{{\hspace{-10 pt} {\large \adforn{9}} \textbf{DeepSense6G Dataset}}} 
To evaluate \textsc{Enwar}'s performance, we utilize a large-scale, real-world multi-modal sensing and communication dataset \cite{deepsense4}. DeepSense6G dataset offers a robust platform for testing \textsc{Enwar}’s ability to interpret complex spatial and environmental data. For our evaluation, we focus on Scenario 36 that includes GPS coordinates of a vehicle equipped with four signal receivers, its captured 360-degree LiDAR point clouds and front-back camera frames, and the GPS coordinates of another vehicle equipped with a transmitter. {We meticulously identified a total of 180 scenes/samples, 30 of which is used for testing,  that captures urban environments with varying numbers of pedestrians, cyclists, and vehicles; an exemplary scene shown in Fig. \ref{fig:sample_answer}} 

\subsubsection*{{\hspace{-10 pt} {\large \adforn{9}} \textbf{Modality Transformation and Information Extraction}}} 

For all the selected scenes, \textsc{Enwar} extracts latitude and longitude coordinates from GPS inputs to determine the positions and relative bearings of two vehicles, which are then converted into textual format for seamless integration into \textsc{Enwar}’s prompt. For front-rear images, \textsc{Enwar} perform image-to-text transformation to generate a textual description of the visual content by using InstructBLIP trained on Vicuna-7b and optimized for visual-tuned instructions \cite{blip}. 

For LiDAR point clouds, \textsc{Enwar} leverages the super fast accurate 3D (SFA3D) model for object detection and analysis [\url{https://github.com/maudzung/Super-Fast-Accurate-3D-Object-Detection}]. SFA3D was modified to extract object information, including locations and bearings relative to the sensor, and converts this data into text to describe the environment. SFA3D utilizes a ResNet-based keypoint feature pyramid network (KFPN) for reliable LiDAR object detection, transforming 3D point clouds into Birds-Eye-View images, which are then processed to identify objects with high confidence, providing detailed information such as positions, dimensions, and orientations. 

{The extracted information from each modality is hard-coded into a template [c.f., white text highlighted with black background in Fig. \ref{fig:sample_answer}] to be utilized during the prompting and grounding process. Since object types and potential blockages are not readily available labels within the DeepSense6G dataset, we manually create ground truth text for all 180 scenes by correcting extracted information if necessary and/or adding missing details. }

\subsubsection*{{\hspace{-10 pt} {\large \adforn{9}} \textbf{Instructional Text Prompt}}} 

{The prompt includes \textit{extracted information} of each scene and specifies predefined tasks, guiding \textsc{Enwar} to accurately analyze the wireless environment, detect potential blockages, and generate relevant insights based on the processed multi-modal inputs[c.f., yellow box in Fig. \ref{fig:sample_answer}].}

\subsubsection*{{\hspace{-10 pt} {\large \adforn{9}}  \textbf{Chunking and Embedding}}}
\textsc{Enwar} utilizes the gte-large-en-v1.5 embedding model from AliBaba-NLP to vectorize ground truth samples for knowledge base creation \cite{embeddings}. This model supports a tokenized context length of up to 8,192 tokens. To maintain continuity between data segments, the transformed textual data is divided into chunks of 1,024 characters, with a 100-character overlap to ensure context is preserved across boundaries.

\subsubsection*{{\hspace{-10 pt} {\large \adforn{9}} \textbf{Knowledge Base Creation and Performance Evaluation}}}
{
\textsc{Enwar} utilizes the Facebook AI Similarity Search (FAISS) library [\url{https://ai.meta.com/tools/faiss/}] to create 7 distinct knowledge bases—one for each modality combination—and its efficient search/retrieval through the top-95\% ranking explained above. \textsc{Enwar}’s performance was thoroughly evaluated by running RAGAS framework across LLMs such as Mistral-7b/8x7b [\url{https://mistral.ai}] and Llama3.1-8/70/405b [\url{https://www.llama.com}], with model sizes ranging from 7 billion to 405 billion parameters. For comparison, baseline versions of these LLMs were used to benchmark the performance of vanilla LLMs against \textsc{Enwar}, particularly in generating detailed and accurate responses.}


\begin{table}[t!]
\centering
\caption{\ac{KPI} comparison for the scene in Fig. \ref{fig:sample_answer}.}
\label{tab:KPI_comp}
\resizebox{\columnwidth}{!}{%
\begin{tabular}{|l|c|c|c|}
\hline
\textbf{KPIs}          & \textbf{Relevancy} & \textbf{Correctness} & \textbf{Faithfulness} \\ \hline
\textbf{Vanilla Llama} & 70.3 \%             & 54.3 \%              & 42.2 \%               \\ \hline
\textbf{\textsc{Enwar}}      & 81.2 \%            & 76.9 \%              & 68.6 \%               \\ \hline
\end{tabular}%
}
\end{table}

\begin{figure*}[t!]
    \centering
    \includegraphics[width= \textwidth] {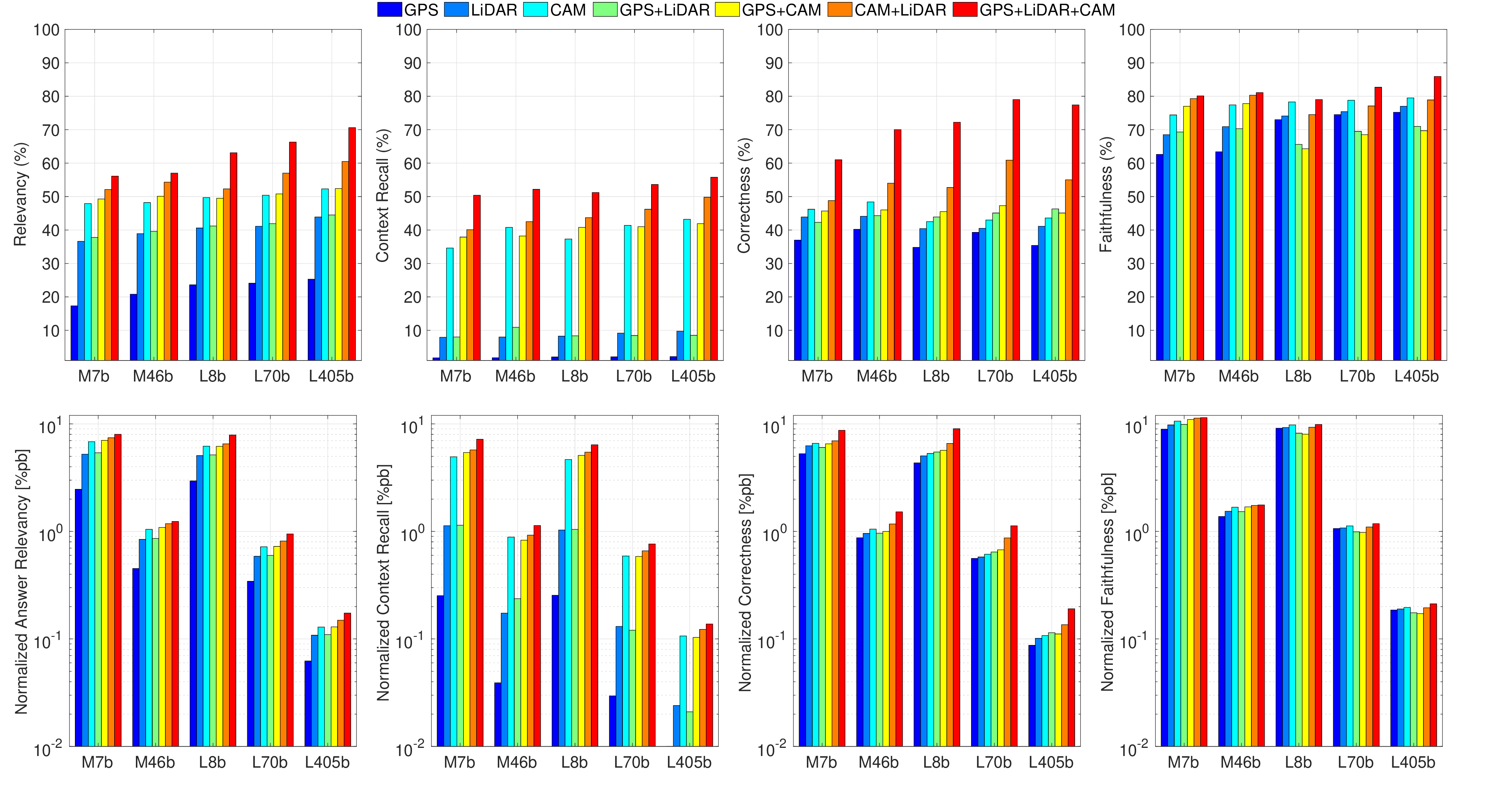}
    \caption{\ac{KPI} comparison of \acp{LLM} across modality combinations: the first and second rows present absolute \ac{KPI} [\%] and \ac{KPI} normalized per billion parameters of each \ac{LLM} [\%pb], respectively.}
    \label{fig:KPI_comparison}
\end{figure*}

\subsection*{\hspace{7pt} \textbf{A Comparative Analysis of an Inference Case Study}}
In this section, we evaluate the perception capabilities of Vanilla Llama and \textsc{Enwar} in processing, analyzing, and interpreting spatial relationships between objects, as well as inferring potential obstacles between two units. As shown in Fig. \ref{fig:sample_answer}, both models were tasked with generating a detailed description of a busy city street scene. To provide more descriptive insights, we selected a specific scenario featuring congested traffic, including cars, motorcycles, pedestrians, and other stationary objects along the road.

While Vanilla Llama provides a general description of the scene and acknowledges the presence of various entities, its response remains superficial. It offers only basic information about the distances and directions of objects, largely reiterating the extracted data without analyzing the relative positions of vehicles or obstacles and failing to infer how these elements might impact movement or communication the units. 

By contrast, as highlighted and underlined in Fig. \ref{fig:sample_answer}, \textsc{Enwar} delivers a detailed, contextually rich breakdown of the spatial dynamics from Unit 1's perspective. It accurately identifies the positions and distances of nearby vehicles, pedestrians, and cyclists, providing a clear analysis of potential obstacles and suggesting maneuvering strategies for Unit 1 in the congested environment. Crucially, it assesses line-of-sight communication between the units and identifies any obstructions, essential for effective coordination. \textsc{Enwar}’s ability to infer network interactions, evaluate line-of-sight, and propose navigation strategies highlights its advanced environment sensing capabilities, setting it apart from Vanilla LLaMa’s more generic output.

At this stage, it is crucial to compare the corresponding \acp{KPI} for the scene depicted in Fig. \ref{fig:sample_answer}. As shown in Table \ref{tab:KPI_comp}, wherein context recall is omitted as it is a \ac{RAG} context-specific metric, \textsc{Enwar} consistently outperforms vanilla Llama in delivering more contextually aligned, accurate, and faithful responses. This underscores its superior ability to interpret complex environments and provide reliable insights, largely due to \textsc{Enwar}’s seamless integration of multi-modal data and its robust \ac{RAG} framework. {\textsc{Enwar}'s single-modality inference times are 100 ms, 100 ms, and 2.5 s for GPS, LiDAR, and camera inputs, respectively. Although image-to-text translation dominates the processing time, it’s important to recognize that scene elements such as traffic conditions, weather, landscape, and area classification (urban, suburban, rural) — as illustrated in Fig. \ref{fig:sample_answer} — typically exhibit limited variability within the order of seconds.  That is visual data updates may not require constant reprocessing at the same frequency as other modalities. Unlike the narrative-driven information extracted from camera inputs, LiDAR and GPS provides quantitative measures within the order of milliseconds, allowing \textsc{Enwar} to utilize windowing for efficient tracking and prediction of environmental changes. These inference times can be further reduced through well-crafted hierarchical LLM architectures as outlined in the final section.}


\section*{\textbf{\ac{KPI} Evaluation of State-of-the-Art LLMs \\ on Modality Combinations}}
This section evaluates the performance of various state-of-the-art \acp{LLM} across different modality combinations, which is presented in Fig. \ref{fig:KPI_comparison} and discussed in the following subsections. 

\subsection*{\hspace{7pt} \textbf{Modality Combination Comparison}}

For single modality evaluations, the general trend shows GPS $<$ LiDAR $<$ CAM in terms of performance across all KPIs. GPS alone provides limited contextual information, resulting in the lowest scores, while CAM proves to be the most effective single modality, offering richer visual context that significantly enhances answer relevancy, correctness, and faithfulness.

When dual modalities are combined, the trends observed in the single-modality evaluations continue. Specifically, when Camera or GPS is paired with LiDAR, CAM+LiDAR consistently outperforms GPS+LiDAR across all KPIs. This reflects the stronger impact of visual data on the models’ ability to generate contextually rich and accurate responses. As expected, the integration of all three modalities yields the highest performance across every KPI. The fusion of spatial, depth, and visual information allows the models to deliver the most comprehensive and accurate responses, further emphasizing the value of multi-modal data integration for advanced situational awareness.

\subsection*{\hspace{7pt} \textbf{\ac{LLM} Type and Size Comparison}}

Across all modality permutations, an increase in the parameter size correlates with improved absolute \ac{KPI} values. Larger models consistently outperform their smaller counterparts across all metrics, reflecting the advantage of parameter scaling in \acp{LLM} performance. However, it is also notable that the rate of \ac{KPI} improvement slows and begins to saturate as the parameter space grows. This indicates diminishing returns at higher parameter counts, with the largest models showing only marginal gains compared to their slightly smaller counterparts. In terms of model comparisons, the performance differences between Mistral 7b and LLaMa 8b are minimal, indicating that these two models are comparable in terms of their effectiveness across KPIs and modality combinations. The second row of Fig. 2 reveals a noticeable observation: Despite the larger models providing better overall absolute \acp{KPI}, the efficiency (i.e., performance per billion parameters) of adding more parameters decreases significantly, potentially indicating overfitting and interesting research directions covered in the next section. Another promising way of inference latency reduction might be training baby LMs to operate directly on the sensory data at the edge to form local \ac{RAG} to eliminate the need for intermediary steps of modality transformation and information extraction. 

\section*{\textbf{Conclusion and Future Directions}}
As a RAG-empowered multi-modal \ac{LLM} framework, \textsc{Enwar} can address some of the key challenges in next-generation networks by enabling situational aware network management through multi-modal perception. By preprocessing and integrating various sensory data, \textsc{Enwar} enhances its ability to interpret complex wireless environments and deliver contextually rich, human-interpretable insights. In spite of promising preliminary results, there is still room for improvement through several architectural enhancements depending on the target applications, which are discussed below.

\subsubsection*{{\hspace{-10 pt} {\large \adforn{72}} \textbf{ Hierarchical and Federated \ac{LLM} Architectures}}} 
For mission-critical and time-sensitive tasks, inference time and model efficiency can be significantly improved by adopting a federated \ac{LLM} architecture that integrates smaller, edge-based "baby LMs" with full-scale \acp{LLM} in the cloud. Baby LMs are designed for near-real-time operation at the edge, reducing reliance on cloud infrastructure. By employing model pruning and quantization techniques, these models remain lightweight and efficient, focusing on immediate, critical tasks. More complex computations are offloaded to cloud-based \acp{LLM}, providing both speed at the edge and scalability in the cloud. Further latency reduction can be achieved by training baby LMs to operate directly on the sensory data at the edge to form local \ac{RAG}, which bypasses intermediary steps of modality transformation and information extraction. 

\subsubsection*{{\hspace{-10 pt} {\large \adforn{72}} \textbf{Serverless \ac{LLM} Architectures}}} 
Serverless architectures allow cloud-based \acp{LLM} to dynamically scale resources based on demand, making them ideal for non-time-sensitive tasks such as data aggregation, post-event analysis, or batch processing. These event-driven systems automatically allocate resources only when required, eliminating idle costs and improving cost-efficiency. Although serverless architectures may introduce minor latency due to cold starts, they are well-suited for applications where real-time processing is not essential. Tasks requiring periodic, large-scale computations can be efficiently managed in the cloud without continuous resource allocation.

\subsubsection*{{\hspace{-10 pt} {\large \adforn{72}} \textbf{Cooperative and Adaptive \ac{RAG} Formation}}} 
 Given the importance of \ac{RAG} in the \textsc{Enwar} framework, a distributed and adaptive \ac{RAG} approach could maintain a global knowledge base, aggregating local knowledge bases across the hierarchical \ac{LLM} structure. This collaborative knowledge system enables baby LMs to efficiently retrieve relevant, up-to-date information without needing to store large amounts of data locally. Adaptive learning techniques further enhance \textsc{Enwar} by continuously refining its understanding of dynamic environments, ensuring effective processing of multi-modal inputs. By dynamically updating the global knowledge base and leveraging adaptive learning, \textsc{Enwar} can balance performance improvements with model efficiency, mitigating the risk of overfitting. This adaptive RAG approach optimizes both knowledge retrieval and system adaptability, allowing local models to respond to real-time data while benefiting from the shared global context. It also mitigates the diminishing returns of scaling large models, optimizes resource usage via serverless computing, and ensures continuous learning for improved performance in complex environments.

\ifCLASSOPTIONcaptionsoff
  \newpage
\fi


\renewenvironment{IEEEbiography}[1]
  {\IEEEbiographynophoto{#1}}
  {\endIEEEbiographynophoto}

\vspace*{-5\baselineskip}
\begin{IEEEbiography}{Ahmed M. Nazar} currently pursues a Ph.D. degree in computer engineering from Iowa State University (ISU), Ames, IA, USA. 
\end{IEEEbiography}
\vspace*{-5\baselineskip}
\begin{IEEEbiography}{Abdulkadir Celik} received a Ph.D. in co-majors of electrical engineering and computer engineering from Iowa State University, Ames, IA, USA, in 2016. He is currently a senior research scientist at KAUST. 
\end{IEEEbiography}
\vspace*{-5\baselineskip}
\begin{IEEEbiography}{Mohamed Y. Selim} received a Ph.D. in computer engineering from Iowa State University, Ames, IA, USA, in 2016; where he is currently an associate teaching professor. 
\end{IEEEbiography}
\vspace*{-5\baselineskip}
\begin{IEEEbiography}{Asmaa Abdallah} received a Ph.D. in electrical engineering from the American University of Beirut, Beirut, Lebanon, in 2020. She is currently a research scientist at KAUST. 
\end{IEEEbiography}
\vspace*{-5\baselineskip}
\begin{IEEEbiography}{Daji Qiao} received a Ph.D. degree in Electrical Engineering from The University of Michigan, Ann Arbor, MI, USA. He is currently a full professor at Iowa State University.
\end{IEEEbiography}
\vspace*{-5\baselineskip}
\begin{IEEEbiography}{Ahmed M. Eltawil} received a Ph.D. degree in electrical engineering from the University of California, Los Angeles, CA, USA, in 2003. He is currently a full professor at KAUST.
\end{IEEEbiography}

\end{document}